\documentclass[
aps,%
12pt,%
final,%
notitlepage,%
oneside,%
onecolumn,%
nobibnotes,%
nofootinbib,%
superscriptaddress,%
noshowpacs,%
centertags]%
{revtex4}

\newcommand{\WI}[2]{#1_{\mathrm{#2}}}

\begin{document}
	\selectlanguage{english}

	\title{Neutron Star Mergers and Gamma-Ray Bursts: Stripping Model}
	
	\author{\firstname{S.}~\surname{Blinnikov}}
	\email{sblinnikov@gmail.com}
	\affiliation{%
		Alikhanov Institute for Theoretical and Experimental Physics, National Research Center ``Kurchatov Institute'', Moscow, Russia
	}%
	\affiliation{%
		Kavli IPMU, Tokyo University, Kashiwa (WPI)
	}
	\author{\framebox{\firstname{D.}~\surname{Nadyozhin}}}
	\email{nadezhin@itep.ru}
	\affiliation{%
		Alikhanov Institute for Theoretical and Experimental Physics, National Research Center ``Kurchatov Institute'', Moscow, Russia
	}%
	\affiliation{%
		National Research Center ``Kurchatov Institute'', Moscow, Russia
	}
	
	\author{\firstname{N.}~\surname{Kramarev}}
	\email{kramarev-nikita@mail.ru}
	\affiliation{%
		Alikhanov Institute for Theoretical and Experimental Physics, National Research Center ``Kurchatov Institute'', Moscow, Russia
	}%
	\affiliation{%
		Moscow State University, Moscow, Russia
	}
	
	\author{\firstname{A.}~\surname{Yudin}}
	\email{yudin@itep.ru}
	\affiliation{%
		Alikhanov Institute for Theoretical and Experimental Physics, National Research Center ``Kurchatov Institute'', Moscow, Russia
	}%
	\affiliation{%
		National Research Center ``Kurchatov Institute'', Moscow, Russia
	}

	
	\begin{abstract}
		This paper provides an overview of the current state of the stripping model for short gamma-ray bursts. After the historical joint detection of the gravitational wave event GW170817 and the accompanying gamma-ray burst GRB170817A, the relation between short gamma-ray bursts and neutron star mergers has been reliably confirmed. We show that many properties of GRB170817A, which turned out to be peculiar in comparison with other short gamma-ray bursts, are naturally explained in the context of the stripping model, specifically, the time (1.7 s) between the peak of the gravitational wave signal and the detection of the gamma-ray burst, its total isotropic energy, and the parameters of the red and blue components of the accompanying kilonova
	\end{abstract}
	
	\maketitle
	
	\section*{INTRODUCTION}
	In terms of detection, cosmic gamma-ray bursts
are radiation flares lasting from fractions of a second
to minutes or even hours. The energies of their radiation
lie in the range from tens keV to GeV. Their population
is divided into two parts: long and short.

The generally accepted idea is that long gamma-ray
bursts are generated during the death of a very massive
star, the core of which collapses to form a black hole.
The accretion process of the surrounding matter can
not only lead to the highly energetic ejection of a significant
part of the star's envelope (the so-called
hypernova), but also to the formation of narrow collimated
ejections of matter (jets). If such a jet is oriented
toward us, it will be detected as a long gamma-ray
burst.
	
Short gamma-ray bursts are thought to form during
a neutron star (NS) merger, or possibly a NS-black
hole merger. This process is usually described using
the merging model, in which two NSs approach each
other through the loss of angular momentum due to
the gravitational wave emission and form a single
object as a result -- a supermassive NS or a black hole.
However, there is an alternative to this mechanism,
which was proposed in \cite{Blinnikov1984}, namely, the stripping
model. Here, the more massive NS strips off and
absorbs the matter of its less massive companion. The
latter explodes upon reaching the lower limit of the NS
mass, which produces a gamma-ray burst. The now
alone and more massive NS (as a result of matter
accretion from its companion, it can, in principle, collapse
into a black hole) leaves the place of interaction
at a significant velocity (up to 1000 km/s).	
	
	Event GW170817 is the sixth event detected by the
LIGO-Virgo gravitational-wave antennas and the first
corresponding to a merger of NSs \cite{Abbott2017a}, not black holes.
The gamma-ray burst GRB170817A was observed 1.7~s
after the signal loss at the GW antennas. This directly
confirmed the connection between short gamma-ray
bursts and NS mergers for the first time. In addition,
this almost simultaneous detection of the GW-event
and the gamma-ray burst, coupled with the known
distance (about 40 Mpc) to the host galaxy NGC4993, made it possible to 
impose restrictions on the
deviation of the gravity propagation velocity $\nu$ from
the speed of light $c$: $|\nu{-}c|/c\lesssim 10^{-15}$ \cite{Abbott2017b}. Eleven hours
later, a visible-light source was also discovered, whose
light curves and spectra correspond to the so-called
``kilonova''~\cite{Metzger2019}. This confirmed that the gamma-ray
burst is accompanied by the synthesis of heavy elements
in the r-process. Thus, the first simultaneous
	observations in gravitational-wave and electromagnetic
channels marked the beginning of a new era of
multi-messenger astronomy~\cite{Margutti2020}.

However, the GRB170817A 
turned out to be peculiar; in particular, it was ten thousand
times weaker than other weak short gamma-ray
bursts with known distances \cite{Abbott2017b}. X-ray and radio
observations do not confirm the presence of a strong
jet either \cite{Dobie2018}. At present, theorists try to artificially
explain these observations by models of a choked jet,
jet cocoon, etc. (see, e.g., \cite{Lazzati2017}, \cite{Gottlieb2018} and \cite{Nakar2018}), where the observation
angle of the jet exceeds $13^\circ$ \cite{Finstad2018}. The optical
observations and the model calculation results of NS
mergers also poorly agree with each other \cite{Siegel2019}. Further,
we will show that many properties of event 170817
are naturally explained by the stripping mechanism, as
opposed to the generally accepted model of an NS
merger.
	
The plan of this paper is as follows: first, we 
describe the characteristic features of the merging and
stripping models for short gamma-ray bursts. Next,
we compare the observational data with the
predictions of both models. Our conclusions are given at the end.

\section*{NS MERGER MODEL}
	Let us consider NSs that form a close binary system.
They approach each other due to the angular
momentum loss of the system for the gravitational
wave emission. The further process is apparently
determined mainly by the masses of the system's components.
If the masses are sufficiently large, on the
order of the solar mass, which is the ``standard'' NS
mass, the merging scenario is realized. At the last
stages of NS merging, a non-conservative mass
exchange takes place, which is caused by two main
processes. In the first process, part of matter is
stripped off the NS surfaces by tidal forces and then
ejected mainly into the merging plane \cite{Freiburghaus1999}. The
ejected cold and dense neutron-excess matter with an
electron fraction $\WI{Y}{e}\lesssim 0.2$ undergoes explosive
decompression \cite{Lattimer1977} followed by r-processes, which
give a long (approximately a week) transient in the
near infrared and optical ranges \cite{Li1998}, later called the
red kilonova \cite{Metzger2019}. The other process is associated with
the fact that immediately upon the contact of NSs, a
part of matter is ``squeezed out'' into the polar regions.
As a result of impact heating, this matter is heated to
high temperatures, which leads to an increase in its
average electron fraction due to weak interactions \cite{Lippuner2015}.
The combined optical and ultraviolet transient generated
by radioactive decays in the matter with
is usually called blue kilonova.\footnote{Note that recently a purple kilonova \cite{Metzger2019} with $\WI{Y}{e}$ from $0.2$  to $0.35$ has also been distinguished} The amount of matter
ejected in a particular process depends on the equation
of state and the NS mass ratio \cite{Bauswein2013}.

Depending on the total mass of the binary system
and equation of the nuclear matter state, the merging
results in a black hole or a rapidly rotating supermassive
NS \cite{Kaplan2014}, which collapses into a black hole in a
time of approximately one second \cite{Murguia2020}, releasing a jet.
The newly formed compact object is surrounded by an
accretion disk: in the course of nonstationary accretion,
a part of the neutron-excess matter spills out, also
contributing to the red \cite{Wu2016,Siegel2018} and blue \cite{Fernandez2013,Just2015} kilonovae.

Note that in almost all the multidimensional
hydrodynamic calculations of the interaction of two
NSs at the late stages of the evolution of the binary system,
which lead precisely to their \emph{merging}, the NS
masses were equal and rather large. Even in a special
study devoted to the case of a large mass ratio of binary
components \cite{Dietrich2017}, the mass of the less massive component
was rather large (on the order of the solar mass).

\section*{STRIPPING MODEL AND LNS EXPLOSION}
	
\begin{figure}[ht]
		\setcaptionmargin{5mm} \onelinecaptionstrue
		\includegraphics[width=0.25\textwidth]{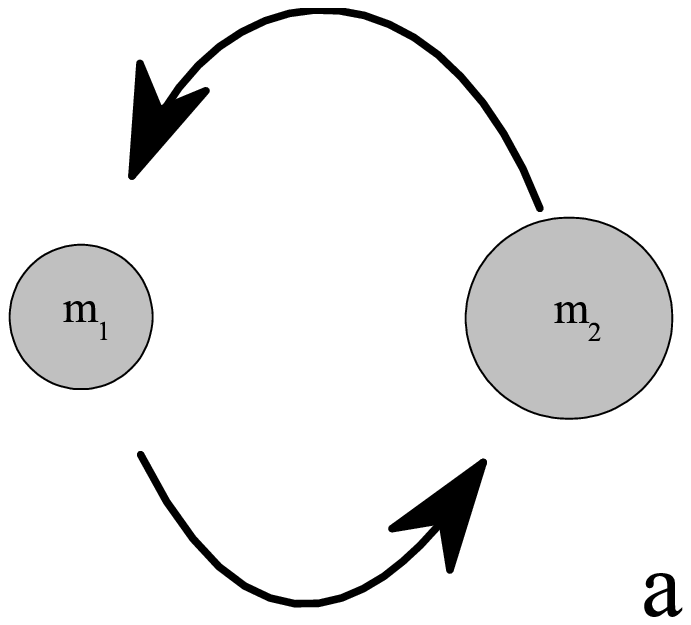}
		\includegraphics[width=0.37\textwidth]{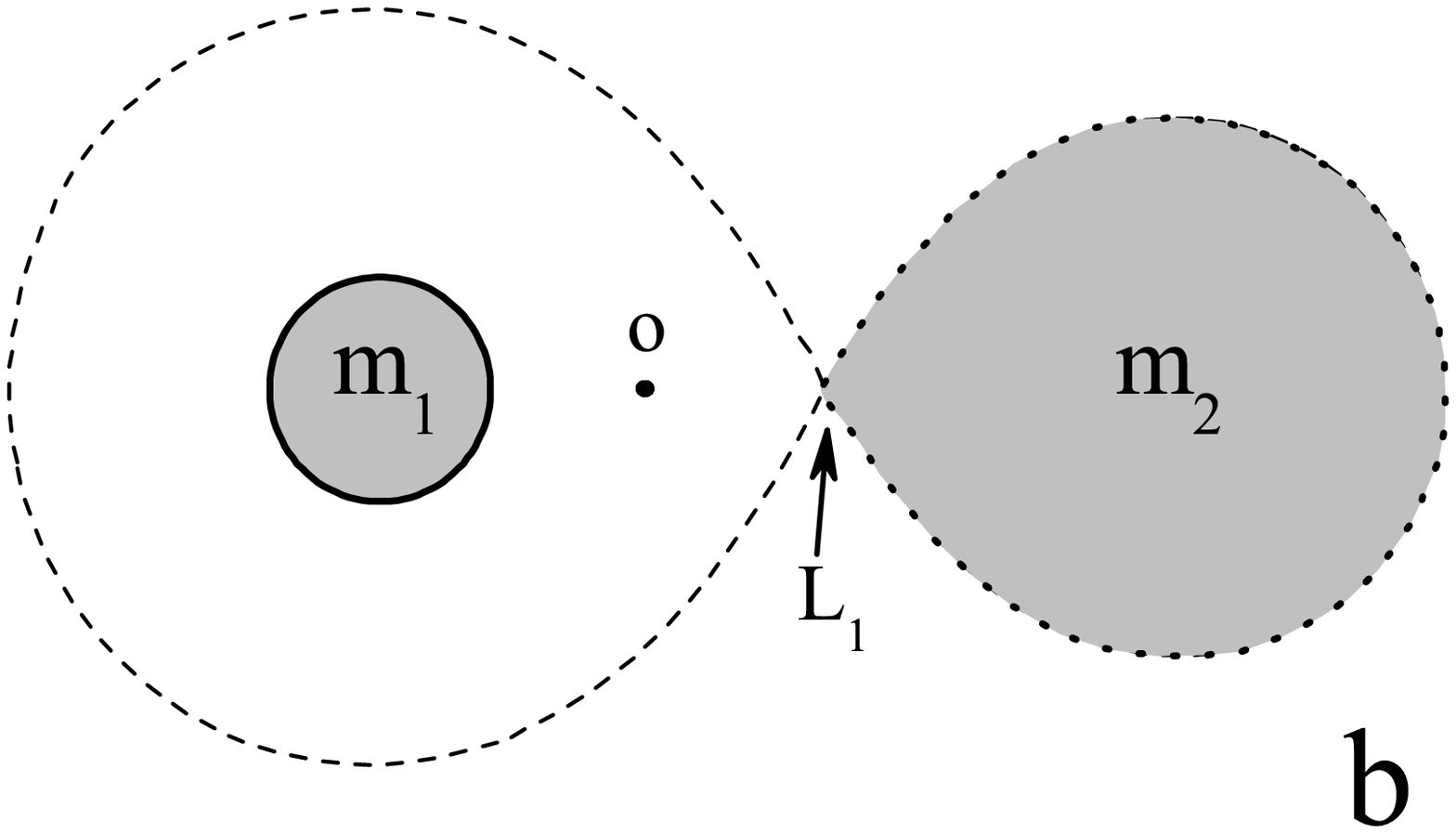}
		\includegraphics[width=0.33\textwidth]{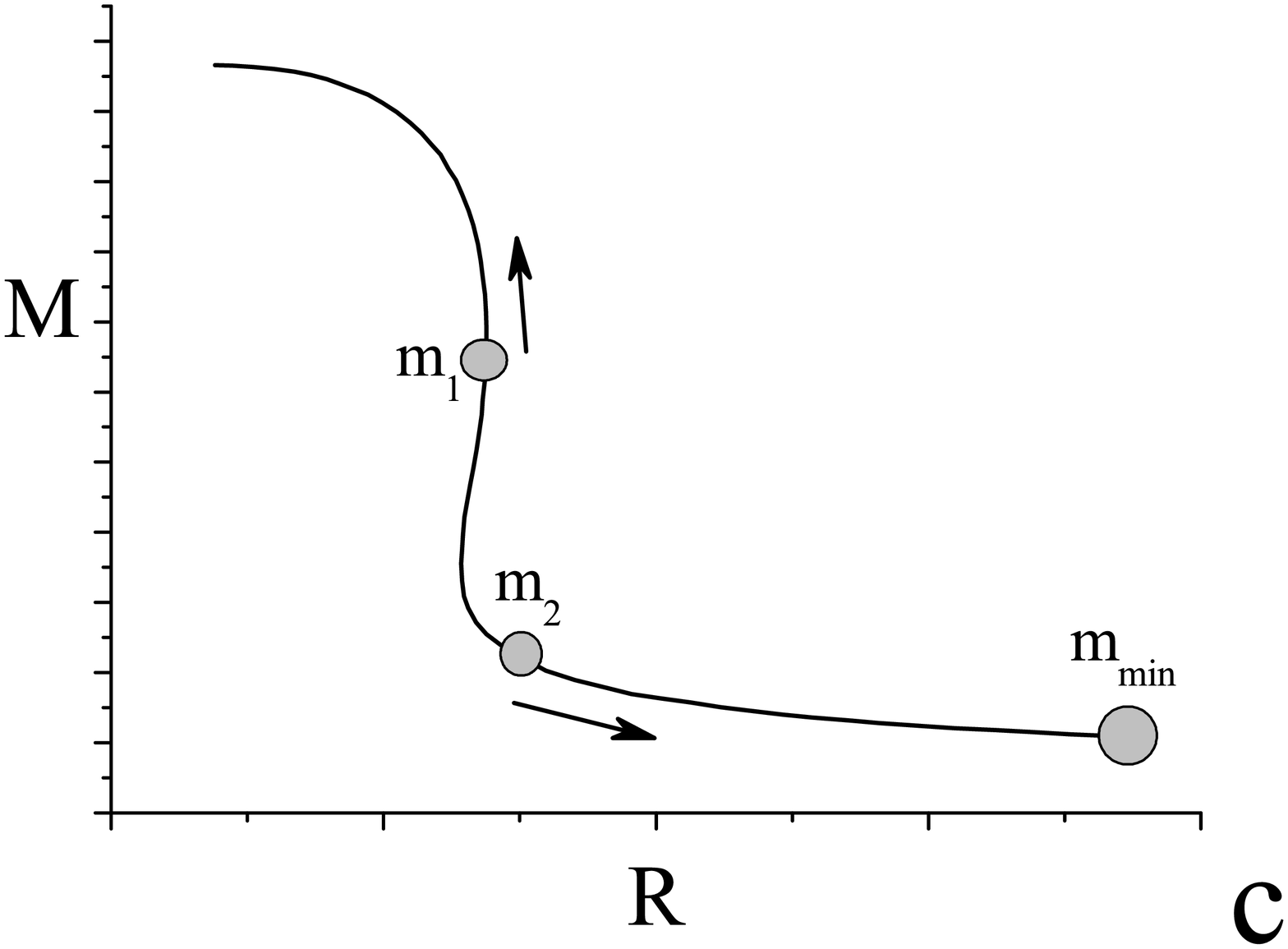}
		\captionstyle{normal}
		\caption{Stripping scenario (schematically): two NSs approach each other due to gravitational emission (a); LNS overfills its Roche
lobe and the flow begins (b); as a result, the binary system components $m_1$ and $m_2$ in the mass-radius diagram move in the directions
indicated by the arrows~(c).}
		\label{Pix:stripping}
\end{figure}

What will change in the scenario described above if
the system is highly asymmetric, i.e., the component
masses differ significantly, and, moreover, the mass of
the low-mass neutron star (LNS) is rather small? Let
us consider the details of the process using Figure~\ref{Pix:stripping}. Fig.~1a shows a binary NS system, in which the component
masses satisfy the condition $m_1>m_2$. At the same
time, the LNS $(m_2)$ has a larger radius. During the
approach, the LNS is the first to overfill its Roche lobe
(see Fig.~1b), and through the inner Lagrange point $L_1$ it begins to flow over onto its massive companion $m_1$. In the mass-radius diagram, the stars begin to
move in the directions indicated by the arrows in
Fig.~1c. For this scenario to be realized, it is important
that the initial LNS mass ($m_2$) was on a shallow branch
of the NS mass-radius curve (Fig.~1c). The specific
value of the LNS mass sufficiently small for the onset
of stripping depends on the equation of the NS matter
state. In \cite{Sotani2014}, fig.~1, a set of NS mass-radius curves in
the low-mass range for various equations of state is
presented. There are significant uncertainties in the
behavior of these curves; however, the characteristic
value of this mass can be roughly estimated as $M\sim 0.5 M_\odot$
	
Let us consider the following aspect of the stripping
scenario: will the process of the matter flow be stable?
Let a part $\triangle m$ of the matter move from $m_2$ to $m_1$. At the
same time, the LNS radius $R_2$ increased (see Fig.~1c).
However, the distance $a$ between the components also
increased since the system became more asymmetric
(a conservative mass exchange is assumed). The effective
size of the LNS Roche lobe $\WI{R}{R}$ has also grown. It
can be parameterized as	
\begin{equation}
\WI{R}{R}=a \WI{f}{R}(q),\quad \WI{f}{R}(q)=0.462\sqrt[3]{\frac{q}{1{+}q}},\label{Roche}
\end{equation}
where $q=m_2/m_1\leq 1$ is the parameter of the asymmetry
of the system. The given approximation for $\WI{f}{R}(q)$ is
the only possible one; see also \cite{Eggleton1983} and \cite{Bisikalo2013}. For the stability
of the matter flow, it is necessary that $\triangle \WI{R}{R}{>}\triangle R_2$.
This brings us to the following condition:	
\begin{equation}
	\left|\frac{d\ln m_2}{d\ln R_2}\right|>\left[2(1{-}q)-(1{+}q)\frac{d\ln \WI{f}{R}(q)}{d\ln q}\right]^{-1}.\label{Roche_staility}
\end{equation}
If we use a specific expression from (\ref{Roche}) for $\WI{f}{R}(q)$, the
expression in square brackets in (\ref{Roche_staility}) will be simplified
to $5/3{-}2q$. Thus, the flow will be stable as long as the
derivative of the LNS mass with respect to its radius
(the absolute value) is sufficiently large. As the star $m_2$
loses mass and shifts to the right along the $(M{-}R)$ diagram
(see Fig.~1c), the dependence becomes
increasingly flat. We used the NS equation of state BSk22
from \cite{Haensel2004}, and the mass of the massive companion was
taken as $m_1=1.4 M_\odot$; it was found that the flow stability
is lost when $m_2\approx 0.107 M_\odot$. In this case, the minimum
NS mass ($\WI{m}{min}$, see Fig.~1c) for this equation of
state is $\WI{m}{min}\approx 0.089 M_\odot$.

Thus, the events in the stripping scenario after the
start of the mass exchange unfold as follows: at first,
the exchange is stable, i.e., the LNS radius increases
more slowly than the critical Roche lobe. The mass
exchange takes place on a long time scale, determined
by the rate at which the system loses its angular
momentum carried away by gravitational radiation.
Only when the LNS reaches a very low mass
($0.107 M_\odot$ in the numerical example above), the stability
of the flow is lost, and the remainder of the $m_2$ matter
is absorbed by $m_1$ on a fast, hydrodynamic time
scale. When $m_2$ reaches the $\WI{m}{min}$ value, i.e., the minimum
NS mass, it loses its hydrodynamic stability and
explodes. This scenario was first calculated in \cite{ClarkEardley1977}.
The electromagnetic radiation burst accompanying
the explosion was proposed by Blinnikov et al. \cite{Blinnikov1984} as a
source of short gamma-ray bursts. In the subsequent
study \cite{Blinnikov1990}, a hydrodynamic calculation of the explosion process of the LNS that reached the minimum
mass was carried out. A comparison of the results with
the observations will be given below. The LNS explosion
was also considered in a number of studies, which
investigated such aspects of the process as the effects of
proper rotation \cite{Yudin2019}, the influence of a massive companion
on the explosion process \cite{Manukovskiy2010}, the accompanying
nucleosynthesis processes \cite{Panov2020}, neutrino radiation
burst \cite{Sumiyoshi1998}, etc. \cite{Colpi1989}. Some historic details of the development
of the stripping scenario can also be found in
\cite{Baklanov2016}.

\section*{COMPARISON WITH OBSERVATIONS}

Let us consider the first stage of the stripping scenario,
following the study by Clark and Eardley \cite{ClarkEardley1977}.
As a numerical example, they chose a system with initial
masses $m_1=1.3~M_\odot$ and $m_2=0.8~M_\odot$. Recall that
the maximum value $m_2$ at which stripping is possible
depends significantly on the equation of state. If we
now compare these masses with the range of masses
derived from the analysis of the gravitational wave
event GW170817 \cite{Abbott2019}, a fairly close agreement will be
found: $m_1\in (1.36\div 1.60)M_\odot$, $m_2\in (1.16\div 1.36) M_\odot$, for
the case of small proper moments of NS rotation, and $m_1\in (1.36\div 1.89)M_\odot$, $m_2\in (1.0\div 1.36) M_\odot$ for the case
of large ones (note that earlier \cite{Abbott2017PRL}, the authors indicated
a wider range $m_2\in (0.86\div1.36) M_\odot$ for the latter
case).

NSs approached each other, and the luminosity of
the gravitational-wave emission continuously
increased until the flow began. After the beginning of
the mass exchange, the stars started to ``move apart'',
the asymmetry of the system increased, and the GW
luminosity began to decrease. The shape of the GW
luminosity curve obtained by Clark and Eardley is
remarkably similar to the LIGO-Virgo observations.
In 1.7 s after the beginning of stripping (which corresponded
to the peak of GW emission), the LNS
reached its minimum mass and exploded. In \cite{Abbott2017b},
fig.~2, an astonishing agreement of the measurement
results with Clark and Eardley's visionary prediction
was demonstrated: after the maximum of the GW
emission curve, the LIGO and Virgo antennas lost
the signal. And then 1.7~s later (!), the FERMI and
INTEGRAL satellites detected a gamma-ray burst.

\begin{figure}[ht]
		\setcaptionmargin{5mm} \onelinecaptionstrue
		\includegraphics[width=0.9\textwidth]{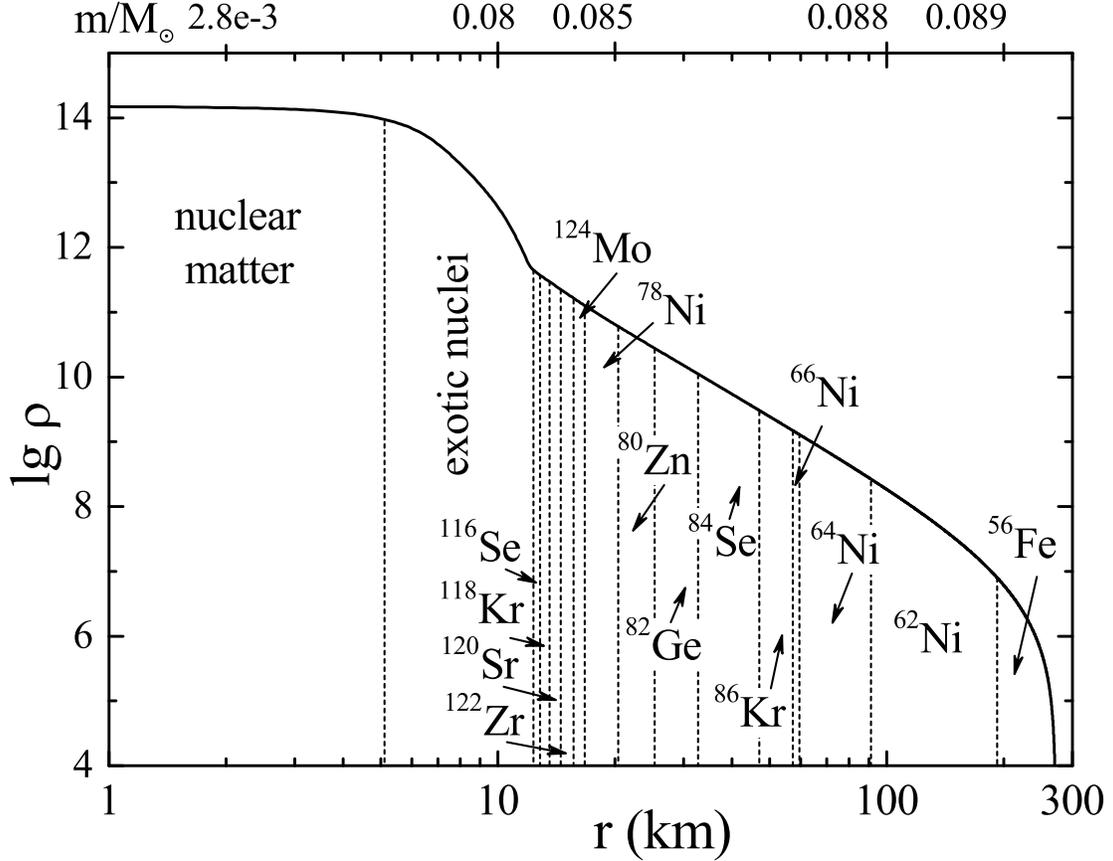}
		\captionstyle{normal}
		\caption{Structure of the minimum-mass LNS. The composition of matter and the dependence of the density logarithm $\lg\rho$ on
the radial coordinate $r$ are shown. The upper abscissa shows the current mass (in solar masses) for several $r$ values.}
		\label{Pix:LowMassNS}
	\end{figure}
Let us now proceed to the key ingredient of the
stripping mechanism, the explosion of the LNS in a
binary system, and consider the structure of the minimum-mass LNS. Figure \ref{Pix:LowMassNS} shows the dependence of its
density logarithm $\lg\rho$ versus the radial coordinate $r$.
The upper abscissa shows the mass values (in solar
masses) for several $r$ values. The structure of matter is
also shown: from the surface inward, the outer crust
consists of increasingly heavy and neutron-excess
nuclei, starting from $^{56}\mathrm{Fe}$ and ending with $^{116}\mathrm{Se}$. The
specific sequence and composition of nuclides may
vary slightly depending on the mass formula and other
parameters of the equation of state for the NS crust
(see, e.g., \cite{Ruster2006}), but the general trend remains the
same. Further comes a layer of exotic nuclear structures
immersed in a sea of emerging free neutrons; at a
density on the order of $\rho\simeq 10^{14}\ \mbox{g}/\mbox{cm}^3$, this layer turns
into a homogeneous nuclear matter. It should be
noted that the entire LNS crust extending over 200~km
contains less than 10\% of the star's total mass. In fact,
the LNS consists of a very dense and small (with a	
	radius of approximately 10~km) core, which contains
nearly the entire mass of the star, and an extended light
envelope.

\begin{figure}[ht]
		\setcaptionmargin{5mm} \onelinecaptionstrue
		\includegraphics[width=0.8\textwidth]{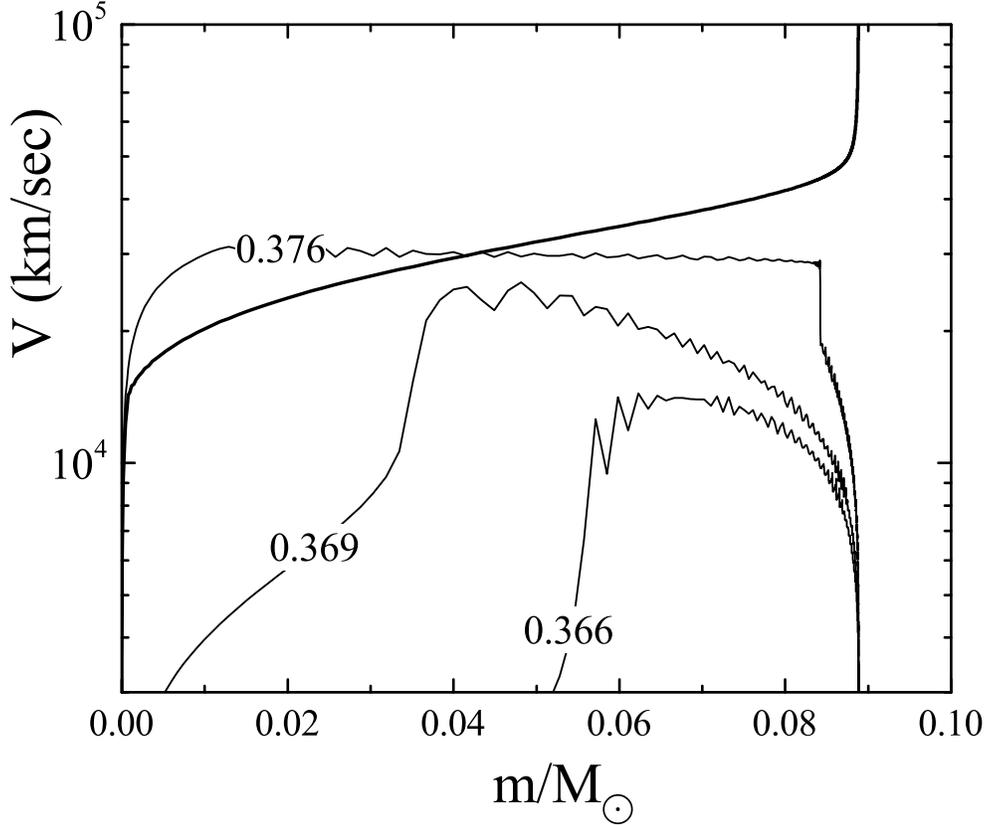}
		\captionstyle{normal}
		\caption{Matter velocity $V$ as a function of mass $m$ in the exploding matter of the LNS. The numbers on the curves show the time
(in seconds) after the loss of stability. The thick line shows the final value of the expansion rate.}
		\label{Pix:Velosity}
	\end{figure}
Let us now consider what happens to the LNS after
the stability loss, following the paper \cite{Blinnikov1990}. Some
details of this process, first calculated by D.~Nadyozhin
in the said study, are shown in Fig.~\ref{Pix:Velosity}. Specifically,
it illustrates the dependence of 
velocity of the LNS matter $V$ (in km/s) as a function
of mass $m$ (in solar masses) inside the star, the so-called
``mass'' coordinate. The numbers on the curves
show the time in seconds after the stability loss. The
thick line shows the final value of the expansion rate
(the velocity of matter at infinity). It can be observed
that the loss of stability and the expansion of matter
starts from the surface and covers the entire star in
approximately a third of a second. Acoustic vibrations
generated at the center propagate along the descending
density profile of the extended LNS shell and
transform into shock waves (see the velocity surge on
the curve at $t=0.376$~s). In this case, the outer layers
are heated to temperatures on the order of $T\sim 10^{9}$~K.
According to the original paper \cite{Blinnikov1990}: ``This should
result in an X-ray and soft gamma ray burst with a total
energy of $10^{43}{-}10^{47}$~erg.'' It was shown in \cite{Abbott2017b}, fig.~4
that the total isotropic energy of the GRB170817A
was more than 3 orders of magnitude
lower than that of other short gamma-ray bursts and
amounted to $3\times 10^{46}$~erg. Here, we also see remarkable
agreement between the stripping model and the
observational data. It is also worth noting that the
LNS shell consisting of various heavy neutron-rich
nuclei (see Fig.~\ref{Pix:LowMassNS}), which is heated by shock waves and
ejected into the surrounding space, is an ideal place for
the r-process \cite{Panov2020}.

\begin{figure}[ht]
		\setcaptionmargin{5mm} \onelinecaptionstrue
		\includegraphics[width=0.8\textwidth]{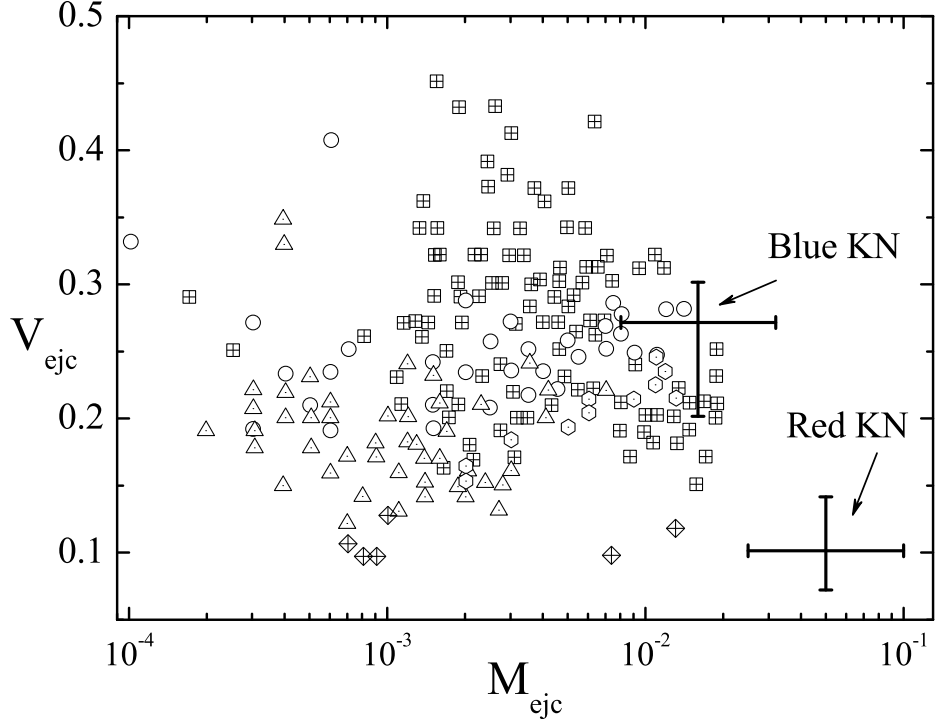}
		\captionstyle{normal}
		\caption{Diagram for ejecta mass $\WI{M}{ej} (M_\odot)$ versus ejecta velocity $\WI{V}{ej}$  (in units of the speed of light) for the blue and red components
of the kilonova. The symbols show the results of numerical calculations in various models of NS merging.}
		\label{Pix:Siegel}
	\end{figure} 
	Let us also refer to the data in Fig.~\ref{Pix:Siegel} adapted from \cite{Siegel2019} (reproduced with the kind consent of the author).
The figure is a diagram for ejecta mass $\WI{M}{ej} (M_\odot)$ versus
ejecta velocity $\WI{V}{ej}$ (in units of the speed of light $c$) for
the matter of the blue and red components of the kilonova,
which have the parameters
	\begin{align}
		&M_\mathrm{ej}^\mathrm{blue}=(1.6^{+1.4}_{-0.8})\times 10^{-2} M_\odot,\quad V_\mathrm{ej}^\mathrm{blue}=(0.27^{+0.03}_{-0.07})c,\label{blue}\\
		&M_\mathrm{ej}^\mathrm{red}=(0.5^{+0.5}_{-0.25})\times 10^{-1} M_\odot,\quad V_\mathrm{ej}^\mathrm{red}=(0.1^{+0.04}_{-0.03})c.\label{red}
	\end{align}
Thus, the blue component has a high velocity (about a
third of the speed of light) and a small mass of the
ejecta, about 1\% of  $M_\odot$, while the red component, on
the contrary, has a low ejecta velocity and a relatively
large mass. The symbols on the same graph show the
modeling results obtained in the merging model by
five different research groups (\cite{Bauswein2013}, \cite{Hotokezaka2013}, \cite{Radice2018}, \cite{Sekiguchi2016} and \cite{Ciolfi2017}). Some of
these models can describe the observed parameters of the blue kilonova. However, none of them explains the
values typical for the red component of the
GRB170817A ejecta.\footnote{It is worth noting that D.~Siegel, author of \cite{Siegel2019} from which we borrowed
Fig.~\ref{Pix:Siegel}, believes that the parameters of the red kilonova can be
explained as the outflow of matter from the accretion disk
around the black hole. However, the calculation considering
weak interactions \cite{Miller2019} disproves this assumption.} Meanwhile, if we turn to our
Fig.~\ref{Pix:Velosity}, we can see that most of the LNS mass (approximately
$0.08 M_\odot$) has velocities on the order of $3\times 10^4~\mbox{km}/\mbox{sec}\sim 0.1c$, and the outermost layers are
accelerated to velocities comparable to the speed of
light, which fully agrees with the observations.

Another important point focuses on the total
kinetic energy of the ejecta. For known short gamma-ray
bursts, it is estimated as $\WI{E}{kin}\sim 10^{49}{-}10^{50}$~erg (see,
e.g., the recent survey \cite{Metzger2019}). Meanwhile, the characteristic
energy for GRB170817A, determined
using parameters (\ref{blue}) and (\ref{red}), is $\WI{E}{kin}\sim 10^{51}$~erg.
However, this is exactly what is given by the stripping
model: according to \cite{Blinnikov1990}, the kinetic energy of the
ejecta during the LNS explosion is $\WI{E}{kin}\approx 9\times 10^{50}$~erg.
The proximity of this energy to the classical energy of	a supernova
 explosion (1 foe = 1 Bethe = $10^{51}$~erg)
once led V.~Imshennik to the formulation of his rotational
mechanism of supernova explosions \cite{Imshennik1992}, in
which the LNS explosion is the most important component.

	However, in \cite{Sumiyoshi1998}, the kinetic energy of the LNS
explosion turned out to be lower: on the order of
$10^{49}$~erg. This could be due to two factors: first, the
authors used the outdated Harrison-Wheeler equation
of state, which gave the minimum mass of a NS
$\WI{M}{min}\approx 0.189~M_\odot$, i.e., almost double the value predicted
by modern equations of state ($\WI{M}{min}\approx 0.089~M_\odot$). From the modern point of view, a NS with
such a large mass has negative total energy and cannot
explode. Second, the authors of \cite{Sumiyoshi1998} considered the
losses due to neutrino emission, although in a greatly
simplified formulation of the problem. This loss ingredient
is indeed absent in our simulations and can
reduce the kinetic energy of the ejection. We are currently
working on preparing an appropriate calculation
that should clarify this aspect of the problem.
	
\section*{CONCLUSIONS}
In summary, the GRB170817A 
associated with the gravitational-wave event
GW170817 confirmed the connection of short
gamma-ray bursts with NS mergers. However, many
of its properties turned out to be unexpected, if considered in the current paradigm, in which two NSs
precisely merge to form a single object. In this case,
not a very large amount of matter should be ejected
from the system, but part of it can form narrow collimated
high-energy jets. Meanwhile, the stripping
mechanism provides a natural explanation to the
entire set of observational data on GRB170817. Here,
we would like to emphasize that one should not make
a choice between the merging and stripping mechanisms.
Most likely, one process takes place under some
conditions, while the other process occurs under others.
For the stripping mechanism to be realized, the
mass of one of the binary system components should
be sufficiently small. However, figuring the specific
value of this threshold mass will require significant
efforts both in refining the equation of state of NS
matter and determining the actual behavior of the NS
mass versus radius curves in the low-mass range, as
well as calculating the process of mass exchange in the
binary NS system, in which one of the components is
an LNS. The proportion of binaries with an LNS
companion among the entire binary NS population is
apparently small. This fraction, which has yet to be
determined, will represent the proportion of the stripping
mechanism of gamma-ray bursts in their general
population. This question is an interesting problem
both for observational astronomy and population synthesis
\cite{Ferdman2020}. For the second observation of the NS
merger, the event GW190425, the accompanying
gamma-ray burst was not detected \cite{AbbottGW190425}. In terms of
the stripping model, this is not surprising: at an estimated
distance of about 160~Mpc, the gamma-ray
burst in our mechanism is beyond the detection limits
(see \citep{Abbott2017b}, fig.~4). On the other hand, the component masses
were larger in this case, and, possibly, there was indeed
a merger either without a noticeable ejection of matter,
or with a jet directed away from us.

		\begin{acknowledgments}
		The authors are grateful to the Russian Foundation for
Basic Research (grant nos. 18-29-21019~mk and 19-52-50014) for the support.	The authors are also grateful to the anonymous reviewer for constructive comments.
	\end{acknowledgments} 
	%
	%
	
	%

\end{document}